\documentstyle[preprint,aps,epsf]{revtex}

\newcommand{\be}{\begin{equation}}
\newcommand{\en}{\end{equation}}

\begin{document}
\draft
\title{Conditions for the importance of inelastic losses in photoemission}
\author{Robert Haslinger$^1$ and Robert Joynt$^2$}
\address{$^1$ Los Alamos National Laboratory, Los Alamos, NM 87545 \\
$^2$ Department of Physics,University of Wisconsin-Madison \\
1150 University Ave,  Madison, WI 53706}
\date{\today}
\maketitle

\begin{abstract}
It is shown that for materials near the metal insulator transition,
inelastic losses can give rise to substantial changes on a
scale of 10-100 meV in the observed lineshape compared to
the intrinsic spectral function of an electron photoemitted from a solid.
These changes arise from the frequency dependence of the loss
function, similar to those directly observed in electron energy loss
experiments.  For good metals and insulators, on the other hand,
the inelastic losses not important.  We derive quantitative conditions
to delineate the various cases.  We find in particular
that La$_{1.2}$Sr$_{1.8}$Mn$_2$O$_7$ does not satisfy
these conditions, so that no significant losses are to be expected
in this material.  This contradicts the reasoning of Schulte {\it et.al}
[Phys. Rev. B {\bf 63}, 165429 (2001)], who, on finding no significant lineshape
effects in this material, concluded that inelastic losses are always unimportant.

\end{abstract}



\section{Introduction}

A recent paper by Schulte {\it et al.} \cite{schulte} treats the subject of
energy loss by an electron photoemitted from a solid. \ Their contention is
''that energy loss processes occurring once the electron is outside the
solid contribute only weakly to the spectrum and can in most cases be either
neglected or treated as a weak structureless background''. \ In particular,
they consider the ohmic loss processes suggested by one of us \cite{joynt}
and find that these ''are only a small contribution to the total spectrum''.
\ They give several arguments to support these statements. \ In this
communication, we consider these arguments. \ We show that, for the
systems of interest in this connection, namely those near the
metal-insulator transition, the ohmic loss processes can, on the contrary,
give a substantial contribution to the spectrum. \ This contribution can
change the interpretation of experimental data. \ This is contrary to the
conclusions of Ref. \cite{schulte}

The conceptual background of the problem is in some ways similar to an
ordinary scattering problem. The photoelectron, after absorbing a photon of
fixed energy, becomes a part of the ''incident beam'' which then scatters
from the system on its way to the detector. \ Unlike the incident beam in an
ordinary well-designed scattering experiment, this beam is not
monochromatic. \ Indeed, it is the energy distribution of this beam that is
of the most experimental interest since it gives the spectrum of the solid.
Any part of the beam which undergoes scattering reflects the properties of the system
only indirectly and we therefore wish to minimize the effects of this
scattering as much as possible. \ Unfortunately, one has little control over
this scattering, and it is poorly understood in most cases \cite{olson}. \
Furthermore, the inelastic mean free path of electrons at 20 eV, a typical
energy for valence-band photoemission (PE), is very short when the electron is in the metal
- on the order of a few nm. \ That the inelastic scattering is very strong
is not surprising since the particle is charged. \

It is nevertheless indubitably true that the observed spectrum often
faithfully reflects the underlying electronic spectrum of the solid. \ For
example, in the case of pure elemental metals such as Na, agreement between
dispersion relations from band structure calculations and the movement of
peaks in angle-resolved photoemission intensities is good. \ This is
due to a combination of factors. \ The relevant energy scales are
in electron volts. \ The electrons that reach the surface of the solid have
either undergone such substantial inelastic scattering that their energies
are well below the chemical potential or they have not suffered any
inelastic scattering. \ In the former case, the electrons may be detected,
but because of their very low energies ($\sim eV$ below the chemical
potential ) they can easily be separated from the signal. \ (Typically,
these ''secondaries'' are very numerous). \ In the latter case, the
electrons are candidates for the desired signal, but they must still get to
the detector without further scattering. \ As long as these additional
inelastic losses do not depend strongly on angle or if
the losses
involve energies that are small compared with the energy scales of interest
in the experiment, then they will not distort the observed spectra in an important way. \

Ref \cite{joynt} is an attempt to show how one may correct for the
additional inelastic scattering after the electron leaves the surface in
valence-band photoemission in cases when such corrections are
important. \ These questions were further developed in Ref.\ \cite{rob}.
In this paper we develop quantitaive criteria to decide when such corrections
are important, at least in the special case that a Drude model describes the
conductivity.

It is crucial to be aware that whether the corrections are important depends on
the resolution of the experiment. \ If resolution is 100 meV or greater, as in classic
band-mapping experiments, then losses on the order of 10 meV
(a more typical scale for ohmic dissipation) are not
important. \ If one is looking for a gap or pseudogap on the order of 20
meV, then relatively small energy losses (that could be ignored in the
1970's) become of paramount importance. \ The point is that these smaller
energy scales can match up with the energy scales associated with the ohmic
losses in metals. \ It is therefore necessary to investigate losses due to
ordinary ohmic dissipation. \ Any theory of this phenomenon must be
consistent with the many historical successes of PE in measuring dispersion
relations and densities of states. \ Hopefully a detailed treatment of the
inelastic scattering will also throw light on situations where theory and
experiment appear currently to be in disagreement.

\section{General Considerations}

In the Born approximation, the probability for electrons to undergo a single
inelastic scattering event losing an energy $\hbar \omega $\ after it has
left the system is
\begin{equation}
P(\omega )=\frac{e^{2}}{\hbar \omega v}%
\mathop{\rm Im}%
\left( \frac{-1}{1+\varepsilon (\omega )}\right) =\frac{4\pi e^{2}}{\hbar
\omega ^{2}v}\,\frac{%
\mathop{\rm Re}%
\sigma (\omega )}{|1+\varepsilon (\omega )|^{2}}  \label{eq:gen}
\end{equation}%
This refers to electrons that are emitted normally at a speed $v$ from an
{\it isotropic} material with dielectric function $\varepsilon (\omega )$
and conductivity $\sigma (\omega ).$ \ In Ref. \cite{joynt} the prefactor
was incorrect. \ The correct prefactor was derived by Mills \cite{mills}. \
As pointed out in Ref.\cite{schulte}, there is a sum rule for this function:%
\begin{equation}
\int_{0}^{\infty }P(\omega )\,\,d\omega =\frac{\pi }{2}\frac{e^{2}}{\hbar v}.
\end{equation}%
For a speed $v$ corresponding to 20 eV, we find $\int_{0}^{\infty }P(\omega
)\,\,d\omega =0.65.$ \ The Born approximation is valid only when the total
scattering probability is small, indicating that we may be near or beyond
the validity of the approximation here. Multiple scattering is possible
or indeed probable.  In view of the complications introduced by the
possibility of multiple
scattering, it is important to examine the assumptions made about it. \ We
look carefully at the consequences of various assumptions.

Assumption (1). Multiple scattering is negligible. \ Unitarity then
implies that the forward scattering probability is 0.35. \ The observed
angle-integrated PE intensity is then%
\begin{equation}
I(\omega )=0.35\,N(\omega )f(\omega )+\int_{\omega }^{\infty }N(\omega
^{\prime })\,f(\omega ^{\prime })\,P(\omega ^{\prime }-\omega )\,d\omega
^{\prime }  \label{eq:int}
\end{equation}%
Here $N(\omega )$ is the density of states that the experiment sets out to
measure. $f(\omega )$ is the Fermi function. \ Evidently, the second term is
comparable in magnitude to the first. \ Let us examine the conditions under
which it can be ignored in experiments, (proceeding to subcases of assumption (1)).
(a) $P(\omega )$ is a very sharply
peaked function about $\omega =0$. \ Then the two terms are the same and $%
N(\omega )$ can be extracted from the data without further ado. \ Obviously,
this will only work if the width $\Delta (\omega )$ of $P(\omega )$ is much
less than the features of the data to be resolved. \ If we call the energy
width of the narrowest feature in the density of states $R$ , then we need $%
\Delta (\omega )<<R$. \ (b) Most of the weight in $P(\omega )$ is at
very high $\omega ,$ say at $\omega >\omega _{\max }.$ \ Let the valence
band, or the part of it of experimental interest, be of energy width $B$. \
In this case, there will be little overlap in the two terms of Eq. \ref%
{eq:int} as long as $\omega _{\max }>B$. \ By focusing on the
appropriate frequency range, we can still extract $N(\omega ).$ \ (c) \ $%
P(\omega )$ is a linear combination of functions satisfying (a) and (b). \

If $P(\omega )$ does not satisfy conditions (a), (b), or (c), then certainly
it cannot be ignored. \ This does not mean that it is impossible to extract $%
N(\omega ),$ merely that substantial deconvolution of the observed intensity
will be required.

Assumption (2). \ The probability of multiple scattering is of order 1.
\ Then the observed intensity is%
\begin{equation}
I(\omega )=P_{0}\,N(\omega )f(\omega )+\int_{\omega }^{\infty }N(\omega
^{\prime })\,f(\omega ^{\prime })\,P(\omega ^{\prime }-\omega )\,d\omega
^{\prime }+\int_{\omega }^{\infty }N(\omega ^{\prime })\,f(\omega ^{\prime
})\,P_{m}(\omega ^{\prime }-\omega )\,d\omega ^{\prime }.
\end{equation}%
$P_{0}$ is the probability of no scattering. $\ P_{m}(\omega )$ is the loss
function for electrons that have scattered more than once. \ The
single-scattering term is constrained by the sum rule, and is always of
order one. \ Unitarity then leads us to expect that $P_{0}$ will be small,
though in the absence of any calculation of the higher-order terms we cannot
say how small. \ Under assumption (2), when can scattering be ignored ? \
Clearly both $P(\omega )$ {\it and} $P_{m}(\omega )$ must satisfy\
conditions 1(a), 1(b), or 1(c). \ Furthermore, we can no longer assume that $%
P_{0}=0.35.$ \ We may still hope to extract $N(\omega ),$ but some
assumption must be made about the form of $P_{m}(\omega )$, and\ $P_{0}$
must be calculated or fit to experiment. \ This is not so hopeless as it
might at first seem. \ If \ $P(\omega )$ satisfies 1(a), (all the single
scattering losses are at low energy), then since the multiple scattering
losses are cumulative, it is also very possible that $P_{m}(\omega )$ will
also satisfy 1(a). \ What is necessary is only that the most probable number
n$_{sc}$ of scattering processes satisfies $n_{sc}\,\Delta (\omega
)<<R .$ \ If, on the other hand, $P(\omega )$ satisfies 1(b), the
scattering is at high energy, then $P_{m}(\omega )$ will automatically also
satisfy 1(b) since subsequent scatterings only increase the energy loss
beyond $\omega _{\max }$ and all losses will exceed $B$. \ Similar remarks
hold if $P(\omega )$ satisfies 1(c). \

In practice, then the main difference between assumptions (1) and (2) is
that $P_{0}\neq 0.35$ and must be treated as a fit parameter, and that some
calculation or assumption must be made about $P_{m}(\omega )$. \ Comparing
assumptions (1) and (2) {\it a priori, }assumption (2) is clearly more
plausible. \ If the probability of a single scattering event is 0.65, the
probability that multiple scattering takes place is surely not expected to
be small. \ This expectation is borne out by experiments in which there are
well-defined modes that can by excited in electron energy loss spectroscopy (EELS). \ In ZnO, for
example, multiquantum losses have been observed corresponding to up to five
phonons \cite{ibach2} Furthemore, if one relaxes the assumption that the
surface is perfect, additional scattering is expected. \

\section{Theory}

In order to make some general quantitiative statements about when ohmic
losses might be important,
we need a model for $\sigma (\omega )$ at low frequencies ($%
\omega \leq 100\,meV)$. The Drude model is often quantitatively
accurate and in many other cases at least provides the best two-parameter fit. \
Exceptional materials that are not well described by this model must be
treated on a case-by-case basis. \ Interband and phonon contributions are
not covered by this model.
\ We have
\begin{equation}
\varepsilon (\omega )=1+\frac{4\pi i\sigma (\omega )}{\omega },
\end{equation}%
and%
\begin{equation}
\sigma (\omega )=\frac{ne^{2}}{m^{\ast }(1/\tau -i\omega )}=\frac{\sigma _{0}%
}{1-i\omega \tau },
\end{equation}%
which leads to a loss function%
\begin{equation}
P(\omega )=\frac{\pi e^{2}\sigma _{0}}{\hbar v\tau ^{2}}\frac{1}{\left(
\omega ^{2}-\omega _{0}^{2}\right) ^{2}+\Gamma ^{4}},
\end{equation}%
where%
\begin{equation}
\omega _{0}=\frac{1}{\sqrt{2}\tau }(4\pi \sigma _{0}\tau -1)^{1/2}
\end{equation}%
and%
\begin{equation}
\Gamma =\frac{1}{\tau }(\frac{1}{4}+\omega _{0}^{2}\tau ^{2})^{1/4}.
\end{equation}%

The loss function has two limiting regimes and we examine these in turn.
The first is when $4\pi \sigma _{0}\tau >>1,$ in which case $\omega _{0}$ is
real. \ The loss function is a Lorentzian in $\omega ^{2}$ with peak at
$\omega _{0}\approx (2\pi \sigma _{0}/\tau )^{1/2}$ \ This is the surface
plasma frequency $\omega _{sp}=\omega _{p}/\sqrt{2},$ where $\omega _{p}$ $%
=(4\pi ne^{2}/m^{\ast })^{1/2}$ is the bulk plasma frequency. \ The width of
$P(\omega)$ in this limit is  $\Delta \approx 1/\tau$.
This regime of the loss
function has been treated in connection with EELS by Mills \cite{mills1}. \ For good metals the plasma
frequency is typically several $eV$, so the weight in the loss function is
almost all at high frequencies. The energy range of interest $B$ is
typically less than $\hbar \omega _{0}$, \ while, to take one example, $4\pi
\sigma _{0}\tau \sim O(10^{5})$ for $Cu$ at $273\,K.$ \ Hence assumption
1(b) is satisfied in good conductors. \ This accounts for the success of
classic PE experiments in metals. \ It is interesting to note that for $Bi$
at $273\,K$, $\ \sigma _{0}\tau \sim O(1).$ \ This is still just within the safe
range, but even elemental semi-metals may be not too far from the regime
where $\omega _{0}$ is imaginary.

The second regime is $4\pi \sigma _{0}\tau <<1,$ in which case $\omega _{0}$
is pure imaginary. \ The peak in the loss function is at zero frequency and
the width\ is $\Delta  \approx 2\pi \sigma _{0}.$ \ When $2\pi \sigma
_{0}<<R$, then condition 1(a) is satisfied, and the inelastic losses are not
important. \ This is the insulating limit, and exemplifies the fact
that an insulator does not absorb electromagnetic energy. \ The sum
rule is satisfied because the loss function passes over to a $\delta -$%
function located at zero energy. \ The condition $2\pi \sigma _{0}<<R $
is rather stringent. \ For $R=10\,meV$, a value now attainable
experimentally, $\sigma _{0}<<2.4\times 10^{-12}s^{-1}$ and $\rho
_{0}=1/\sigma _{0}>>375\,m\Omega -cm.$ \ In their paper, Schulte {\it et
al.} appear to make the assumption that the inelastic losses are a
monotonic function of the resistivity - that the losses should be more
severe as a material approaches the insulating limit. \ This is unwarranted.
\ An insulator surely cannot have dissipation at low frequencies.

It is in between these two limits, where $4\pi \sigma _{0}\tau \approx 1$, that
the losses will tend to obscure $N(\omega )$ the most.  In
this crossover regime $P(\omega)$ is spread out over a range of frequencies
between $\omega=0$ and $\omega= (2\pi \sigma _{0}/\tau )^{1/2}$.
Such a behavior of $P(\omega)$ has the potential to cause
large modifications in the observed density
of states.  Figures of $P(\omega)$ and $N(\omega)$ in these various regimes
are given in Ref. \cite{rob}.

The above results allow us to locate the crossover
regime in the $\sigma_0 - \tau$ plane and put numerical bounds on it.  In the
insulating regime ($4\pi\sigma_0\tau < 1$), the peak is at $\omega=0$ and its width
scales as $\Delta = 2\pi\sigma_0$.  In order for the losses to
be observable, $\Delta$ must be wider than R, the resolution of the
detector.  On the other hand, if $P(\omega)$
is too wide then the losses at any one frequency will be so small
that no loss will be observed.  Let us
assume that if $P(\omega)$ is spread out over frequencies greater than the
total energy range of interest B then the losses will be too small for
observation.  These considerations give us one set of bounds.
\be
\frac{R}{2\pi} < \sigma_0 < \frac{B}{2\pi} \qquad \qquad (4\pi\sigma_0\tau<1)
\en
In the metallic limit ($4\pi\sigma_0\tau >> 1$)
$P(\omega)$ is peaked at $\omega_0 =(4\pi \sigma _{0}/2\tau )^{1/2}$.
Obviously if $\omega_0 <R$ then the losses
all located at frequencies less than the resolution of the experiment.
If on the other hand, the peak is at $\omega_0 > B$ the losses are also not of
interest.  Thus for the region $4\pi\sigma_0\tau > 1$ we have the
different set of bounds:
\be
\frac{R^2}{2\pi} < \frac{\sigma_0}{\tau} = \frac{\omega_p^2}{4\pi}
< \frac{B^2}{2\pi}  \qquad \qquad (4\pi\sigma_0\tau<1)
\en
It should also be noted that in the metallic limit $P(\omega)$ is
narrower than R for $\tau>1/R$ and hence may in this case be described by a
delta function located at $\omega_0$.

\begin{figure}[tb]
\epsfxsize \columnwidth
\epsffile{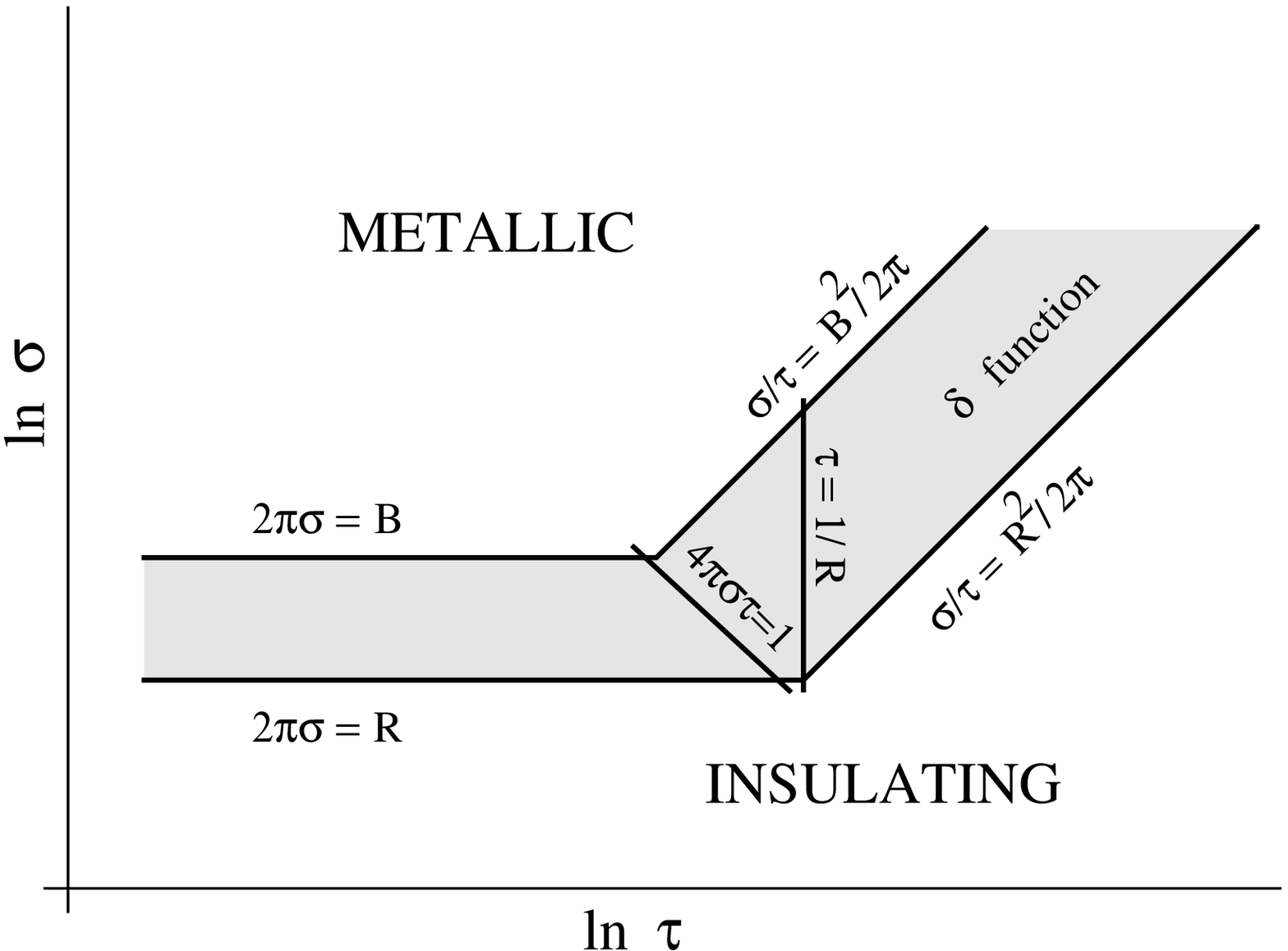}
\caption{Schematic diagram to illustrate when ohmic losses
are important for photoemission studies of cubic materials described by a Drude model.
$\sigma_0$ is the DC conductivity and $\tau$ is the relaxation time.
R is the resolution of the
detector or the size of the smallest feature of interest.  B is the bandwidth
or the total energy range of interest.  Ohmic losses are not important in
either the metallic or insulating regime.  They are potentially important
in the shaded region in between these two limits.  Within the shaded region
the loss function can be characterized as follows.  To the left of the
$4\pi\sigma_0\tau$ line $P(\omega)$ is peaked about zero frequency.  To the
right it is peaked about $\omega_0 = \sqrt{2\pi\sigma_0/\tau}$.  To the
right of the $\tau=1/R$ line $P(\omega)$ may be approximated by a delta function
at $\omega_0$.}
\label{fig:bounds}
\end{figure}

We plot these bounds in figure \ref{fig:bounds}.  The crossover regime comprises
a swath several orders of magnitude wide cutting through the $\sigma_0 -\tau$ plane.
While outside the crossover regime (in either the metalic or insulating regime)
ohmic losses may be neglected this is not the case in the crossover regime.
Detailed calculation using Eq. \ref{eq:gen} is of course necessary to determine the exact
characteristics of any loss since we are considering systems near the metal-insulator
transition and the Drude model does not necessarily apply.
Still, knowledge of $\sigma_0$ and $\omega_p$ should
allow one to make a "rule of thumb" estimate as to whether ohmic loss
is important for a particular material, in which case more detailed calculations
may be in order.  The regime where losses are important
$4\pi \sigma _{0}\tau \approx 1,$ can
also be written as $\omega _{p}\approx \sigma _{0}\approx 1$ $eV.$ \ In
ordinary units, this is $\rho _{0}=1/\sigma _{0}\approx 0.6$\thinspace $
m\Omega -cm,$ close to the Mott value.  The figure shows clearly that
materials that undergo a metal-insulator transition necessarily pass through
the "dangerous" regime.


\section{\protect\bigskip Experiment}

The experiments presented by Schulte {\it et al. }are on the material La$%
_{1.2}$Sr$_{1.8}$Mn$_{2}$O$_{7}.$  This material is highly anisotropic
as their own experiments on the resistivity show. (Fig. 3 of Ref. \cite{schulte})
$\sigma _{zz}\approx 0.1\,\sigma _{xx}$ over the whole
range of temperatures measured. \ The theory described above and in Refs.
\cite{joynt},\cite{rob} applies only to electrically isotropic materials
$(\sigma _{xx}=\sigma _{yy}=\sigma _{zz})$ as was explicitly stated in those
references. \ Formula \ref{eq:gen} would not be expected to apply even
approximately to this material. \ Nevertheless, these authors have
apparently used this theory to analyze their data, as judged from their
discussion of the relation between Figs. 4 and 5 of Ref. \cite{schulte}. \
Although our theory does not apply to this material we can make some
general statements.
La$_{1.2}$Sr$_{1.8}$Mn$_{2}$O$_{7}.$ is not even close to being a Drude
conductor as can be seen in the optical conductivity measurements of
Ishikawa {\it et. al.} \cite{ishikawa} Both the in-plane and out-of-plane $\sigma_1(\omega)$
are very small for $\omega$ close to zero, peaking around 500 meV for the
in-plane conductivity and around 4.5 eV for the out-of-plane.
The absence of a low energy component to both the in- and out-of-
plane conductivities places
this material squarely in the insulating regime.  Low frequency ohmic
losses are not likely here.

In their analysis, Schulte {\it et al.} utilized reflection EELS spectra
to derive the loss function $P(\omega)$ which they then applied to a
constant density of states.  The claim was made that the experimental
data could only be fit by using $P_0$ as a fit parameter, a reasonable
approach as was argued above, and then choosing a very small value for $P_0$.
However, the "zero loss peak" in
Schulte {\it et al.}'s EELS spectra appears to be over 100 meV wide
as is shown Figure 4 of Ref. \cite{schulte}.   Ohmic losses are on the order
of 10-100 meV as was stated above.  The subraction of a "zero loss peak"
100 meV in width will certainly obscure any ohmic
loss effects.  In short, the energy scales used in Schulte {\it et al.}'s
analysis are too large to make any statements about ohmic loss.

La$_{1.2}$Sr$_{1.8}$Mn$_{2}$O$_{7}$ is a tetragonal material, and
any description of inelastic losses in non-cubic systems is
much more complicated than that based on Eq. 1.  In addition, it is
not a Drude conductor and hence the analyses in this
paper do not apply numerically.  Still, if we allow
ourselves to make order-of-magnitude estimates based on the considerations
of this paper, it appears likely that very significant ohmic losses
are not to be expected in La$_{1.2}$Sr$_{1.8}$Mn$_{2}$O$_{7}$, and that
the observed gap is intrinsic.

\section{Conclusion}

As our results have shown, ohmic loss is potentially important
for systems near the metal-insulator transition.  They are not
important for good metals or good insulators.
This accounts for the many successes of photoemission since ohmic losses
may be disregarded in those limits.  The situation is different
in the crossover region and care should be taken in these
materials when interpreting features on the 10-100 meV scale.

We agree with Schulte {\it et.al.} that ohmic losses
are probably not important in La$_{1.2}$Sr$_{1.8}$Mn$_{2}$O$_{7}$
since this material most likely lies in the insulating regime.
However, this material is not a good candidate for testing
the theory of inelastic processes in PE since its resistivity is extremely
high.  Furthermore, it is tetragonal and the theory as it presently
stands only to cubic crystals.  Studies to determine the effect of
ohmic losses on both tetragonal and orthorhombic materials
are currently underway.

\bigskip

\bigskip

\bigskip

\bigskip This work is supported by the NSF under the Materials Theory
program, Grant No. DMR-0081039. (R.H and R.J), by DR Project 200153 (R.H.)
and by the Department of
Energy, under contract W-7405-ENG-36. (R.H.)

\end{document}